**Behavioural/Systems/Cognitive**

# Affine invariance of human hand movements: a direct test

Abbreviated title : Direct test of affine invariance

Quang-Cuong Pham[1], Daniel Bennequin[2]

[1] Laboratoire de Physiologie de la Perception et de l'Action, Collège de France CNRS UMR 7152, 11 place Marcelin Berthelot, 75005 Paris, France

[2] Équipe Géométrie et Dynamique, Institut de Mathématiques de Jussieu, UMR 7586, Paris, France

3 Figures, 1 Table

23 pages

Abstract : 141 words

Introduction : 723 words

Discussion : 1286 words

Total number of words (Abstract, Introduction, Materials & Methods, Results, Discussion, References and Figure Legends) : 4500 words

**Acknowledgments**

We would like to deeply thank Alain Berthoz, Ronit Fuchs and Tamar Flash for stimulating discussions. Daniel Bennequin was supported in part by the European project RoboSoM.

# Abstract

Geometrical invariance, in particular affine invariance, has been recently proposed as an important principle underlying the production of hand movements. However, tests of affine invariance have traditionally been applied to the consequences of this principle rather to the principle itself. Here, we designed and performed an original, direct, test of affine invariance in a scribbling experiment. In each of the 10800 pairs of randomly-selected scribbling segments, we compared the time parameterizations obtained by transforming the first segment using four different transportation rules – affine, equi-affine, Euclidian and constant – with the experimentally-observed parameterization of the second segment. We observed that, when the two paths are affinely-similar, the affine transportation of the first segment yields the time parameterization that best matches the experimental parameterization of the second segment, which directly demonstrates the existence of affine invariance in the production of hand movements.

# Introduction

A fundamental problem in Neuroscience is to understand how human movements are planned and executed. One proposed framework to tackle this problem involves the hypothesis that the motor system may act as an optimal (feedback and feedforward) controller, which might minimize energy expenses, movement variance (Harris and Wolpert 1998), movement duration (Harris and Wolpert 2006), or maximize smoothness, information, accuracy, etc. (see Engelbrecht 2001 for a review). Researchers have also investigated how the motor system performs probabilistic computations in presence of uncertainty and noise (Todorov 2004) and how it adapts to biophysical constraints. However, to the best of our knowledge, no framework has been able to fully explain several universal and well documented properties of human movements. Such properties include the isochrony principle (i.e. the tendency to perform movements of different scales, e.g. drawing a short and a long straight segment, in similar durations, see Binet and Courtier 1893, Lacquaniti et al 1983) and the compositionality principle (i.e. that trajectories tend to be segmented in smaller and simpler standard pieces, Viviani and Cenzato 1985). The existence of these universal properties advocates for a central command by the brain on the planning and execution of movements (cf. Bernstein 1967).

Recently, Bennequin et al (2010) suggested that a *geometrical invariance principle* can complete optimality principles to explain these properties. In particular, the empirical two-thirds power law, relating curvature and velocity in drawing trajectories (Viviani and Terzuelo 1982), could be explained by geometrical invariance (Flash and Handzel 1996, Pollick and Sapiro 1997, Flash and Handzel 2007).

More precisely, the notion of invariance with respect to a group of transformations $G$ (for instance the group of affine transformations acting in the horizontal plane) can be formulated as follows: given two trajectories $A$ and $B$ (a trajectory is a geometric path endowed with a time parameterization on this path), if the path of $B$ can be deduced from that of $A$ by a transformation $g$ belonging to *G*, then the time parameterization of *B* is also deduced from that of *A* by applying *g*.

This general definition yields a number of consequences. For instance, assuming the invariance of movements with respect to a given group *G* prescribes certain relationships between the shapes of the paths and their time parameterizations. Bennequin et al (2010) tested such relationships for the the Euclidian group (length preserving transformations), the equi-affine group (area preserving transformations) and the full affine group (which includes not only dilatations and displacements, but also shearings and stretchings). They showed that these three invariances are present in the production of movements, in various and precisely determined proportions depending on the global shape of the trajectories and the context (for instance, full affine and equi-affine invariances underlie the production of hand movements while equi-affine and Euclidian invariances underlie locomotion). This supports the principle of computational parsimony: once a path is given, the time parameterization can be deduced following these proportions.

The invariance principle might also underlie the choice of the shapes of the paths. For instance, there might exist a preference for paths which are auto-invariant with respect to a given group *G*, like parabolic paths for the equi-affine group (Polyakov et al 2009). Again, this would be related to a parsimony principle: such auto-invariant paths would require less changes in the motor commands.

One can note that the tests presented in the previous studies of geometric invariance (Pollick and Sapiro 1997, Flash and Handzel 2007, Bennequin et al 2010) are concerned with the *consequences* of movement invariance rather than with the invariance itself. For instance, to test the existence of full invariance with respect to a group *G* in a given movement, one usually predicts a velocity profile based on the movement path and *assuming* G-invariance. One then compares this predicted velocity profile with that experimentally observed. While such tests provide a wealth of information, they are indirect and cannot therefore establish with certainty the existence of *G*-invariance.

We propose here an original test, which examines *directly* the general definition of invariance. In particular, our test does not assume *a priori* that the movement follows any invariance. More precisely, we tested whether, when two random trajectory paths happen to be similar through a given affine transformation, their time parameterizations are also similar through this same transformation. For comparison, we also examined invariances with respect to equi-affine and Euclidian transformations.

# Methods

## Subjects, materials, protocol

Four subjects, one male and three females, participated in this experiment. All subjects were right-handed. A white A3 (29.7cm×42.0cm) sheet of paper was placed horizontally on a table. In each trial, a subject sat in front of the table and had to make scribblings on the sheet's surface with his right index. A marker was attached to the tip of his index. The 3D position of this marker was recorded with the Vicon® (Oxford Metrics, Oxford, UK) infrared camera system at 100Hz frequency. The subject was asked to make continuous movements with his index while respecting the following three constraints: stay within the sheet, fill the sheet equally, avoid abrupt movements and cusps [see a typical scribbling in Fig. 1(A,B)]. Each trial consisted of 30s of scribbling. The subject made three trials (separated by ~1 minute) at his preferred speed. He was then asked to make three other trials, at a faster speed. A total of 24 trajectories (4 subjects × 6 trials) were thus recorded in this experiment.

## Analysis

### Comparison of paths, velocity profiles and trajectories

In this study, a *path* designates a geometric object devoid of any time parameterization while a *trajectory* (or trajectory segment) designates a pair: path + velocity profile along that path.

To assess the similarity of two *paths* $\Lambda_1$ and $\Lambda_2$, we used the Hausdorff distance which is defined by

$$d_H(\Lambda_1, \Lambda_2) = \max \{\sup_{x \in \Lambda_1} \inf_{y \in \Lambda_2} \|x-y\|, \sup_{y \in \Lambda_2} \inf_{x \in \Lambda_1} \|x-y\|\} \quad (1)$$

The similarity of two *trajectories* $\Gamma_1$ and $\Gamma_2$ defined over $[0,T]$ was evaluated by the $\mathscr{L}_2$ distance

$$d_{L_2}(\Gamma_1, \Gamma_2) = \frac{1}{T}\sqrt{\int_0^T \|\Gamma_1(t) - \Gamma_2(t)\|^2 \, dt} \quad (2)$$

By extension, we also call this distance a distance between two "timings".

To compare the velocity profiles, we first normalized them so that their average values throughout the movement equals 1 (see Methods of Pham et al 2007). We then assessed the similarity of two normalized velocity profiles by

$$d_v(v_1, v_2) = \sqrt{\int_0^1 (v_1(t) - v_2(t))^2 \, dt} \qquad (3)$$

*Extraction of trajectory segments*

From each recorded trajectory, we extracted 40 random trajectory segments. The duration $T_i$ of each segment was selected uniformly randomly in the interval $(T_{min}, 3T_{min})$ where $T_{min}=0.1$s for the trajectory with largest average speed. This minimum duration was chosen in agreement with previous findings (Llinas 1991, Schnitzler et al 2006). In each trial, $T_{min}$ was chosen such that the average lengths of the segments equal that of the fastest trajectory just mentioned. Overall, the average duration of the segments was 0.53±0.38s. The starting time of each sub-trajectory was also selected randomly from a uniform distribution. Fig. 1(A) shows two random segments extracted from a given trajectory.

Given a sub-trajectory $\Gamma_i = (p_i(t))_{t \in [0, T_i]}$ indexed by $i$, we note $\Lambda_i$ its geometric path, $(v_i(t))_{t \in [0, T_i]}$ its time-varying velocity vector and $L_i$ its length. Out of the 780 possible pairs of segments of each trial, we extracted 450 random pairs. Thus, the database for each speed condition consisted of 5400=12×450 pairs of segments.

*Optimal affine transformation*

Given two segments $\Gamma_1$ and $\Gamma_2$, we looked first for an affine transformation fa which minimizes the Hausdorff distance between the geometric path of fa ($\Gamma_1$) and that of $\Gamma_2$ [see Fig. 1(C) for an example]. This was done by numerically optimizing the six coefficients that define any affine transformation (four corresponding to the linear map and two corresponding to the translation).

*Affine, equi-affine, Euclidian and constant parameterizations*

Since $f_a$ is optimal with respect to the Hausdorff distance, we know that the geometric path $\Lambda$ of $f_a(\Gamma_1)$ is "close" to $\Lambda_2$. We endowed $\Lambda$ with four different parameterizations, corresponding to four different ways of "transporting" the temporal structure of $\Gamma_1$ onto $\Lambda$. These parameterizations can be formulated equivalently under the form of velocity profiles.

The velocity profile corresponding to the *affine transportation*, denoted $v_A$, was simply defined as the velocity profile of $f_a(\Gamma_1)$, i.e. for all $t \in [0, T_1]$

$$v_A(t) = \|\frac{d f_a(\boldsymbol{p}_1)}{dt}\| = \|\boldsymbol{J}_{f_a} \boldsymbol{v}_1(t)\| = \|A_a \boldsymbol{v}_1(t)\| \tag{4}$$

where $\boldsymbol{J}_f$ denotes the Jacobian matrix of a mapping $f$ and $A_a$ denotes the linear mapping associated with the affine transformation $f_a$. Thus, if hand trajectories are perfectly *affine-invariant* and that $\Lambda$ perfectly matches $\Lambda_2$, then we will have $T_1 = T_2$ and $v_A = v_2$ for all $t \in [0, T_1]$.

The velocity profile corresponding to the *equi-affine transportation*, denoted $v_{EA}$, was computed using the equi-affine velocities. More precisely, we first computed $\tilde{v}_1$, the equi-affine velocity profile of $\Gamma_1$ by for all $t \in [0, T_1]$

$$\tilde{v}_1(t) = \frac{v_1(t)}{R(t)^{1/3}} \tag{5}$$

where $R(t)$ is the instantaneous radius of curvature of $\Gamma_1$ at time $t$. We then parameterized $\Lambda$ such that $\tilde{v}$, the resulting equi-affine velocity on $\Lambda$, verifies

$$\tilde{v}(t) = \frac{\tilde{L}}{\tilde{L}_1} \tilde{v}_1(t) \tag{6}$$

where $\tilde{L}$ and $\tilde{L}_1$ are the equi-affine lengths of $\Lambda$ and $\Lambda_1$ (the rescaling was done to ensure that we have $\int_0^{T_1} \tilde{v}(t) dt = \tilde{L}$). Finally, the (Euclidian) velocity profile $v_{EA}$ can be easily computed back from $\tilde{v}$. Thus, if hand trajectories are perfectly *Euclidian-invariant* and that $\Lambda$ perfectly matches $\Lambda_2$, then we will have $T_1 = \tilde{L}_1/\tilde{L}_2 \, T_2$ and $\tilde{v}(t) = \frac{\tilde{L}_2}{\tilde{L}_1} \tilde{v}_2(t)$ for all $t \in [0, T_1]$, where $\tilde{L}_2$ is the equi-affine length and $\tilde{v}_2$ is the equi-affine velocity profile of $\Gamma_2$.

The velocity profile corresponding to the *Euclidian transportation*, denoted $v_E$, was obtained by

defining for all $t \in [0, T_1]$

$$v_E(t) = \frac{L}{L_1} v_1(t) \tag{7}$$

where L is the length of $\Lambda$ (the rescaling was done to ensure that we have $\int_0^{T_1} v_E(t) \, dt = L$ ). Thus, if hand trajectories are perfectly Euclidian-invariant and that $\Lambda$ perfectly matches $\Lambda_2$, then we will have $T_1 = L_1/L_2 \cdot T_2$ and $v_E(t) = L_2/L_1 \cdot v_2(t)$ for all $t \in [0, T_1]$.

Finally, as a "control" parameterization, we considered the *constant* velocity profile $v_C$, which was defined by, for all $t \in [0, T_1]$

$$v_C(t) = \frac{L}{T_1} \tag{8}$$

# Results

The average speed was 0.32±0.11ms$^{-1}$ over the 12 normal speed trials and 0.74±0.20ms$^{-1}$ over the 12 fast speed trials.

### Testing affine invariance

We first tested directly the prediction mentioned in Bennequin et al (2010) and recalled in the Introduction: *"when curve segments are similar under transformations belonging to the (affine) group, the parameterizations of these segments will also be similar"*. Here, the similarity of two "curve segments" $\Lambda_1$ and $\Lambda_2$ under transformations of the affine group can be measured by the Hausdorff distance $d_H(\Lambda,\Lambda_2)$ between $\Lambda_2$ and the best affine transformation $\Lambda$ of $\Lambda_1$. The similarity of the parameterizations can be measured by the distance $d_v(v_A,v_2)$ between $v_A$, the velocity profile on $\Lambda$ defined by the *affine transportation* from $\Gamma_1$, and $v_2$, the velocity profile of $\Gamma_2$. Thus, in statistical terms, the prediction we mentioned implies that there exists a positive correlation between $d_H(\Lambda,\Lambda_2)$ and $d_v(v_A,v_2)$.

Fig. 2(A,E) show indeed a clear positive correlation between $d_H(\Lambda,\Lambda_2)$ and $d_v(v_A,v_2)$ in both speed conditions. Also, a positive correlation existed between $d_H(\Lambda,\Lambda_2)$ and $d_v(v_{EA},v_2)$ where $v_{EA}$ is the velocity profile associated with the *equi-affine transportation* from $\Gamma_1$ [Fig. 2(B,F)]

These positive correlations are not trivial since, by contrast, there was practically no correlation between $d_H(\Lambda,\Lambda_2)$ and $d_v(v_E,v_2)$ [Fig. 2(C,G)] or between $d_H(\Lambda,\Lambda_2)$ [Fig. 2(D,H)] and $d_v(v_C,v_2)$ where $v_E$ and $v_C$ are respectively the velocity profile associated with the *Euclidian transportation* from $\Gamma_1$ and the constant velocity profile.

Qualitatively, one can observe in Fig. 1(D) that $v_A$ (green line) and $v_{EA}$ (dashed magenta line) were more similar to the velocity profile of $\Gamma_2$ (blue line) than the initial velocity profile of $\Gamma_1$ (red line) or the constant velocity profile (dashed cyan line).

### Comparing different parameterizations

It could be argued that the positive correlation observed previously resulted solely from the fact that $v_A$ and $v_{EA}$ yielded bad results for large Hausdorff distances. We thus concentrate here on the pairs of segments with low Hausdorff distances (that is, whose paths were affinely similar). In addition, to avoid artifacts caused by the normalization of the velocities, we examined directly the distance between trajectories (or between "timings", see Methods), that is, the $\mathscr{L}_2$ distance between $\Gamma_1$ and $\Lambda$, where the latter was endowed respectively with $v_A$, $v_{EA}$, $v_E$ and $v_C$.

We first categorized the pairs into bins according to their Hausdorff distances expressed in percentages of the length of $\Gamma_2$. The size of each bin is given in Table 1. For each bin, we computed the average and the SD of the $\mathscr{L}_2$ distances across the pairs in the bin.

The results were plotted in Fig. 3(A,B). One can clearly observe that, in both speed conditions, the affine transportation yielded the lowest $\mathscr{L}_2$ distances, followed by the equi-affine, the constant and the Euclidian transportations (in that order) for low Hausdorff distances (up to $d_H \approx 2.5\%$ of $L_2$ in the normal speed condition and up to $d_H \approx 2\%$ of $L_2$ in the fast speed condition).

In the normal speed condition, considering the 792 pairs with $1\% \leq d_H \leq 2.5\%$, the average $\mathscr{L}_2$ distance was 4.5% for the affine transportation, 5.0% for the equi-affine transportation, 5.2% for the constant transportation and 6.2% for the Euclidian transportation. The difference between the "affine" and the "equi-affine" means was significant (Wilcoxon paired test, $p=7.8 \cdot 10^{-29}$), that between the "equi-affine" and the "constant" means was not significant (Wilcoxon paired test, $p=0.17$), and that between the "constant" and the "Euclidian" was significant (Wilcoxon paired test, $p=5.6 \cdot 10^{-21}$).

In the fast speed condition, considering the 165 pairs with $1\% \leq d_H \leq 2\%$, the average $\mathscr{L}_2$ distance was 3.5% for the affine transportation, 4.3% for the equi-affine transportation, 4.5% for the constant transportation and 5.2% for the Euclidian transportation. The difference between the "affine" and the "equi-affine" means was significant (Wilcoxon paired test, $p=1.0 \cdot 10^{-6}$), that between the "equi-affine" and the "constant" means was not significant (Wilcoxon paired test, $p=0.81$), and that between the "constant" and the "Euclidian" was significant (Wilcoxon paired test, $p=2.8 \cdot 10^{-4}$).

Finally, it could be argued that the good results of the parameterizations for small Hausdorff distances may come from the fact that the transformations of this category are purely Euclidian. To verify this, we computed, for each affine transformation $f$, the following "pseudo-norm"

$$n(f) = \|A\| + \|A^{-1}\| \qquad (9)$$

where A is the matrix associated with $f$ and $\|A\| = \sqrt{\text{trace}(AA^T)}$. We have $n(f) \geq 2\sqrt{2} \simeq 2.8$ with equality if and only if $f$ is Euclidian. For instance, $n(f) \simeq 3.6$ if $A = \begin{pmatrix} 1 & 0 \\ 0 & 2 \end{pmatrix}$. For each bin, we then computed the $n(f)$ across the transformations of each bin, after thresholding the outliers [if $n(f) > 20$ then set $n(f) = 20$], which were less than 9% of each bin.

The results showed that the transformations considered were not purely Euclidian (average n(f)>4 in each bin) and no clear dependance of the average n(f) on the Hausdorff distances was observed [Fig. 3(C)].

# Discussion

We have designed and performed a new, direct, test of geometrical invariance for hand movements. From a set of scribbling data, we chose 5400 random pairs of trajectory segments in each of two speed conditions. For each pair, we compared the parameterization obtained from the first segment by four different transportation rules with the experimentally-observed parameterization of the second segment. The first rule was affine: the parameterization was defined by the affine transformation that best superposes the two geometrical paths. The second and third rules were equi-affine and Euclidian, transporting the linear equi-affine and Euclidian velocity patterns respectively. The last rule simply imposed a constant velocity profile, irrespective of the initial velocity pattern.

Our test is more direct than the tests presented in Bennequin et al (2010) and complementary to them. In particular, our test is a purely *relative* test, which assumes no a priori kinematic laws (see Introduction). Also note that, since the total movement durations were normalized in this study, this is not a test of isochrony as that performed by e.g. Lacquaniti et al (1983), see also below.

The results showed a positive correlation, in the two speed conditions, between (a) the Hausdorff distance between the second segment with the image of the first by the best affine transformation, and (b) the distance between the affine parameterization and the experimental one. This correlation did not exist for the Euclidian transportation or for the constant velocity, implying the non-triviality of the result.

The correlation for the equi-affine transportation was positive, with a bigger slope than that corresponding to the affine transportation. However, in both speed conditions, the affine transportation performed better than the equi-affine transportation for small Hausdorff distances, that is, for the most affinely-similar segments. This last observation also implies that the two-thirds law (which follows from equi-affine invariance) is not a sufficient explanation.

An element that may have contributed to this comparatively inferior performance of the equi-affine transportation could be related to the necessity of calculating the Euclidian curvature in the process of computing the equi-affine parameterization (see equation 5). While the computation of all the other parameterizations requires a first-order derivative, the computation of the equi-affine parameterization thus implies to differentiate twice. Given the limited the number of points of the shortest segments (a 0.1s segment corresponds to 10 sample points), this generated a certain rugosity of the equi-affine velocity profile, which could have increased the distance between this profile and any smooth profile, and thus could have penalized the equi-affine transportation.

Another element that should be taken into account in the comparison between the equi-affine and affine transportations is that the latter takes into account isochrony (Bennequin et al 2010). However, in the present analysis, we normalized the velocity profiles, such that this comparative advantage of affine invariance disappears. Once this difference is suppressed, equi-affine and affine transportations become quite similar. For instance over second-order curves (conics), which give local approximation at order four of any trajectory, it is not possible to distinguish normalized equi-affine from affine parameterization. Moreover it has been shown that the paths produced spontaneously by human or monkey subjects are well approximated by segments of conics (Sosnik et al 2004, Polyakov et al 2009). Thus, it is not surprising that the performances of equi-affine and affine transportations were quite similar.

A more detailed study for discriminating between equi-affine and affine transportations should take in account the size of the dilatation induced by the transformation (their determinant).

Finally, it would be interesting to perform a similar study for locomotion, in particular to assess directly the importance of Euclidian invariance in the context of locomotion, since we know from Bennequin et al (2010) that Euclidian invariance may play a more important role in locomotion than in hand drawing.

More generally, we believe that the methodology we developed for studying directly geometrical invariances can become an essential element in the toolkit of movement and perception neuroscientists.

**The functional role of affine invariance**

The role of geometrical invariance for perception has been highlighted for a long time by e.g. Helmoltz in 1876 and Poincaré in 1902 (see the references in Bennequin et al 2010). Piaget's work (1956) emphasized the role of geometrical invariance – from topology to projective, affine and Euclidian geometries – in development and learning. More recently, this line of thought was further developed by Koenderink and Van Doorn (1991). Also, Todd et al (2001) insisted more particularly on the pertinence of affine geometry in perception. As perception and action are deeply intertwined (cf. Berthoz 1991), it is natural to expect that geometrical invariance – and particularly affine invariance, which takes its roots in force applications and changes of perspective – also plays a large role in the *production* of movements.

In particular, one would expect that movement timing reflects the possible symmetries of trajectories and their possible transformations during execution. In fact, a direct relation between space and time could be a basic computational principle of motor control. Such a relation is offered by geometric invariance principles.

Conceptually, we suggest that the general framework for invariance is a kind of "dual information theory". This point of view is based on the Galois' approach for resolving algebraic equations, and extended by Lie and Cartan for solving systems of partial differential equations or geometrical comparison problems. In the context of motor control, this would correspond to the following hypothesis: a large ambiguity on future movements is reflected in the Central Nervous System by a large symmetry group (acting on the states of activity of neurons or on motor commands). This would be similar, in the case of perception, to the fact that a large uncertainty is reflected in the sensory system by a probabilistic distribution with high entropy. Each new step in the execution, resulting from choices already made or from new decisions based on more recent knowledge, would reduce the ambiguity group in smaller ones, in the same way that new sensory inputs reduce the entropy of the uncertainty structure. Correlatively, a trajectory can be interpreted (through Cartan's frame correspondence, see Cartan 1937 and Bennequin et al 2010) as a description of the space around the subject, in a similar way that new sensory inputs increase the mutual information between the sensory system and the external world. In this context, it is natural to expect, in movement production, the use of the largest possible invariance (in particular, affine invariance), which is dual to the information maximization principle in perception.

As explained by Bennequin et al (2010), the affine group is the largest possible one for a physically reasonable invariance theory of planar movements. However they also experimentally observed that smaller subgroups of this group appeared naturally, such as the Euclidian group and the equi-affine group. Thus, one of the main consequences of Bennequin et al (2010)'s results was the necessity to use *mixtures* of kinematical laws based on each of these geometries. An example of such mixtures is the deformation of the affine law by the Euclidian law when the trajectory curvature decreases. Here, we only considered pure invariances, but it is probable that *mixtures* of standard invariances could yield even better results in our tests, in a similar way that, in the probabilistic theory of perception, probability laws appear as mixtures of uniform laws on subsets of data.

In conclusion, we suggest that geometrical invariance should constitute a part of a general theory of movement production, along with stochastic control and dynamical stochastic computations. Such a general theory should also be compatible with optimality considerations. Developing a general framework that combines and organizes the notions of invariance, stochasticity and optimality to explain the production of movements thus represents a challenging but fundamental task for movement neuroscientists.

# Figures' legends

### Figure 1

Typical trajectories and trajectory segments. (**A**) The geometric path of a typical trajectory. Two random segments are highlighted in blue and red. (**B**) The velocity profile of the entire trajectory. The velocity profiles of the random segments of (A) are highlighted in the same colors. (**C**) Geometric paths of the segments $\Gamma_1$ (red), $\Gamma_2$ (blue) and of the image of $\Gamma_1$ by the optimal affine transformation [$f_a(\Gamma_1)$, green]. (**D**) Normalized velocity profiles of $\Gamma_1$ (red), $\Gamma_2$ (blue) and $f_a(\Gamma_1)$ (green). The normalized velocity profile associated with the equi-affine transportation (dotted magenta line) and the constant normalized velocity profile (dotted cyan line) are also depicted.

### Figure 2

Correlation between spatial similarity and parameterization similarity. Each pair of segments was represented by a dot whose X-coordinate was the Hausdorff distance between the two trajectories of the pair and whose Y-coordinate was the distance between the two velocity profiles. (**A,B,C,D**) velocity profiles associated with the affine, equi-affine, Euclidian and constant transportations, normal speed. (**E,F,G,H**) velocity profiles associated with the affine, equi-affine, Euclidian and constant transportations, fast speed.

### Figure 3

Comparison of different parameterizations. Average and SD of the $\mathscr{L}_2$ distance between $\Gamma_2$ and $\Lambda$ endowed with respectively the affine (green), equi-affine (magenta), Euclidian (red) and constant (cyan) velocity profiles. (**A**) normal speed. (**B**) fast speed. (**C**) Average "pseudo-norm" (for the definition, see Results) of the affine transformations in each bin in the normal speed (black) and fast speed (gray) conditions.

# Table

| Hausdorff distance (in %) | Normal speed | Fast speed |
|---|---|---|
| $0 < d_H \leq 1.00$ | 7 (discarded) | 9 (discarded) |
| $1.00 < d_H \leq 1.25$ | 44 | 8 |
| $1.25 < d_H \leq 1.50$ | 78 | 22 |
| $1.50 < d_H \leq 1.75$ | 143 | 57 |
| $1.75 < d_H \leq 2.00$ | 193 | 78 |
| $2.00 < d_H \leq 2.25$ | 174 | 135 |
| $2.25 < d_H \leq 2.50$ | 160 | 152 |
| $2.50 < d_H \leq 2.75$ | 125 | 174 |
| $2.75 < d_H \leq 3.00$ | 112 | 209 |
| $3.00 < d_H \leq 3.25$ | 143 | 230 |
| $3.25 < d_H \leq 3.50$ | 143 | 270 |
| $3.50 < d_H \leq 3.75$ | 172 | 226 |
| $3.75 < d_H \leq 4.00$ | 152 | 231 |
| $4.00 < d_H \leq 4.25$ | 158 | 222 |
| $4.25 < d_H \leq 4.50$ | 157 | 186 |
| $4.50 < d_H \leq 4.75$ | 172 | 206 |
| $4.75 < d_H \leq 5.00$ | 172 | 201 |

**Table 1** Number of pairs of trajectory segments in each bin (cf Fig. 3).

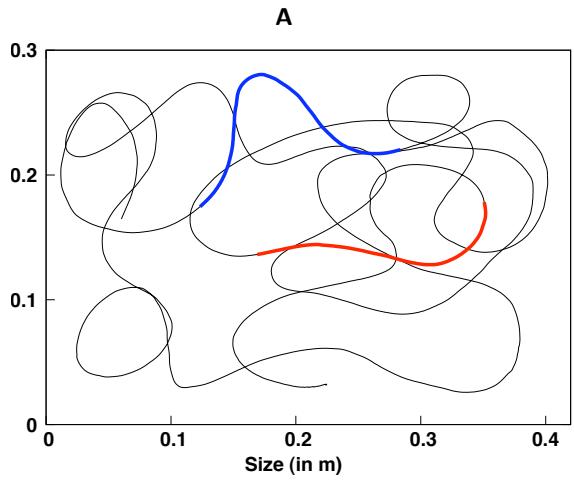
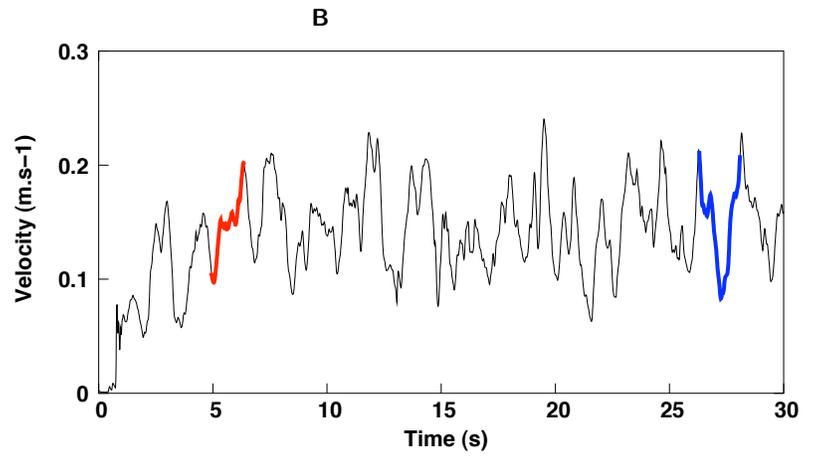
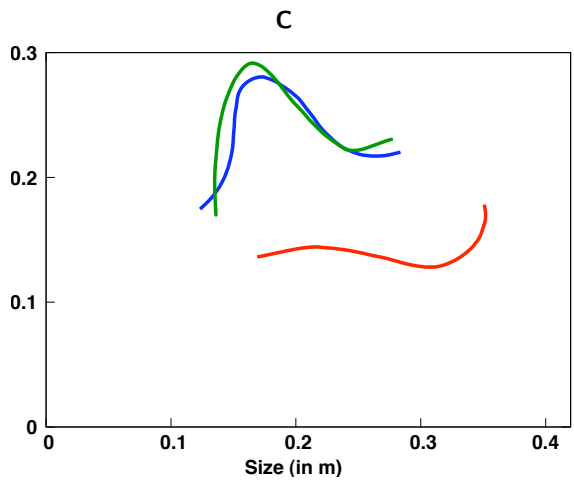
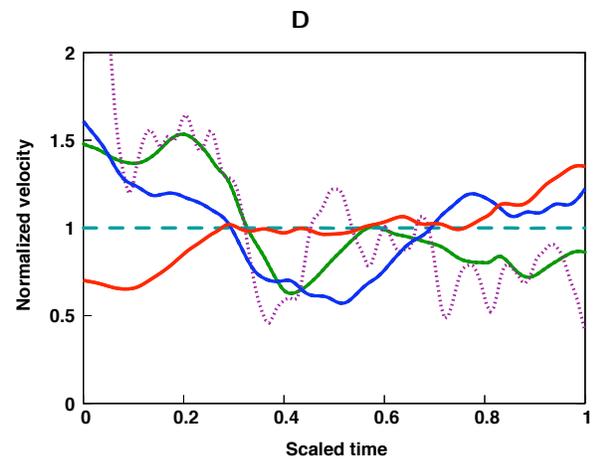

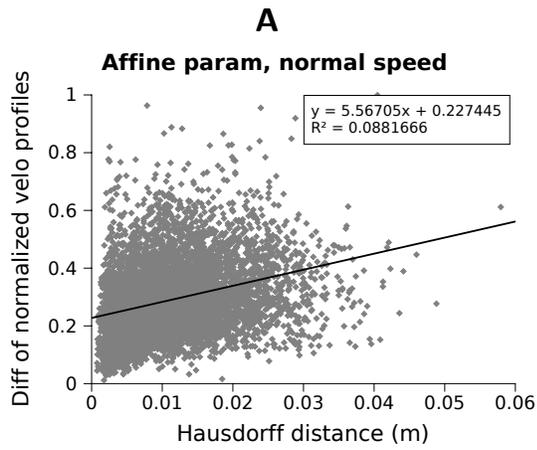
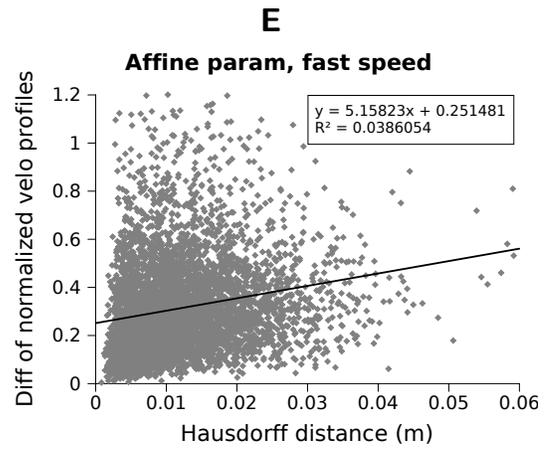
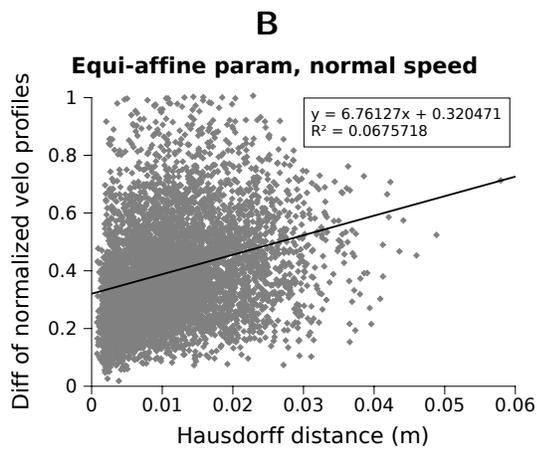
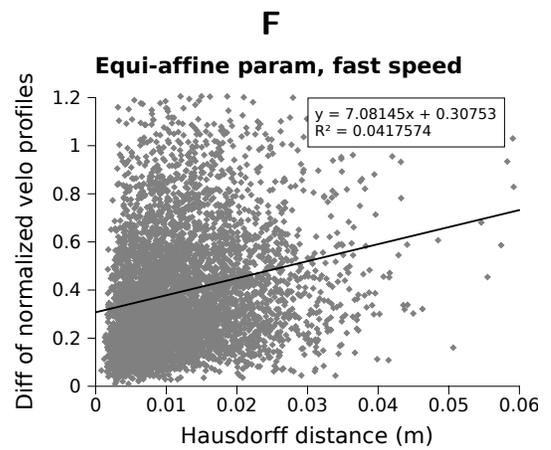
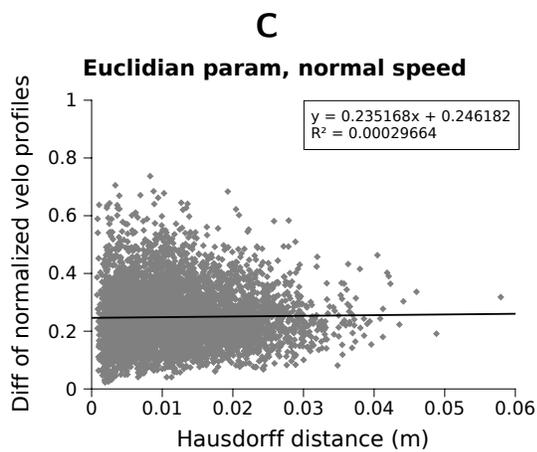
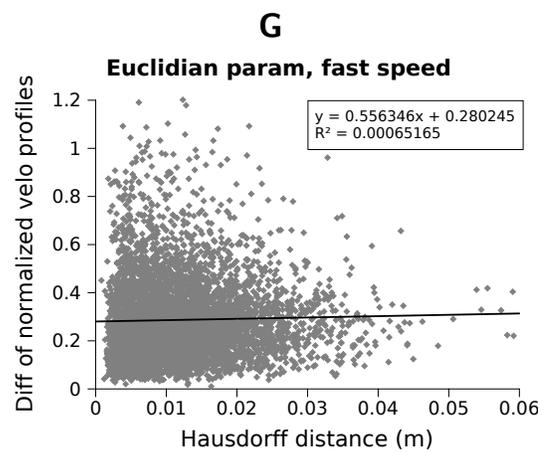
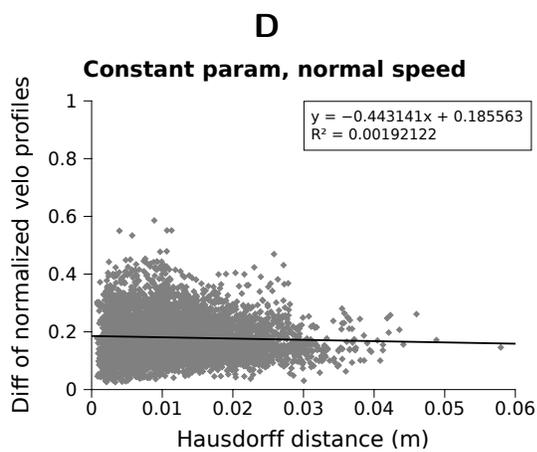
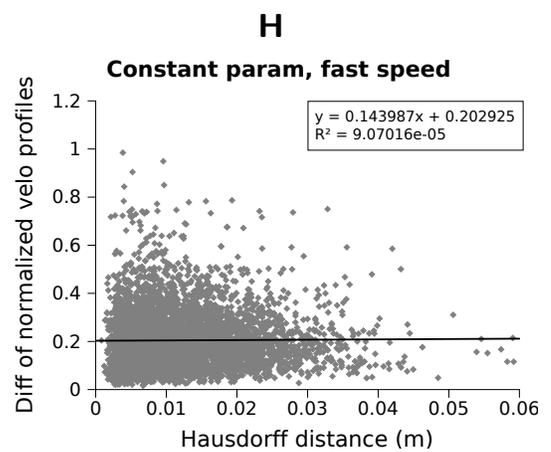

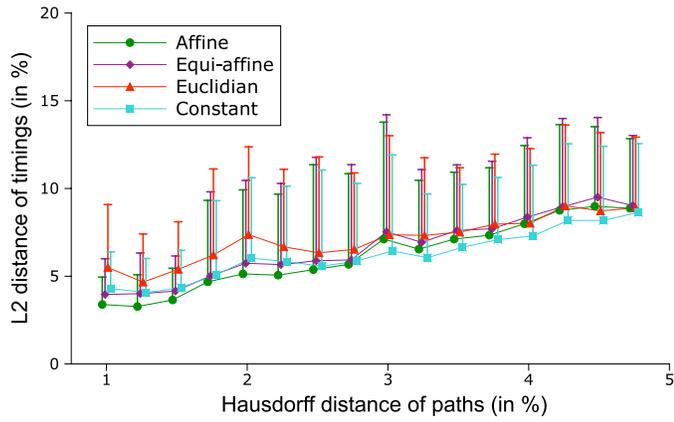

A

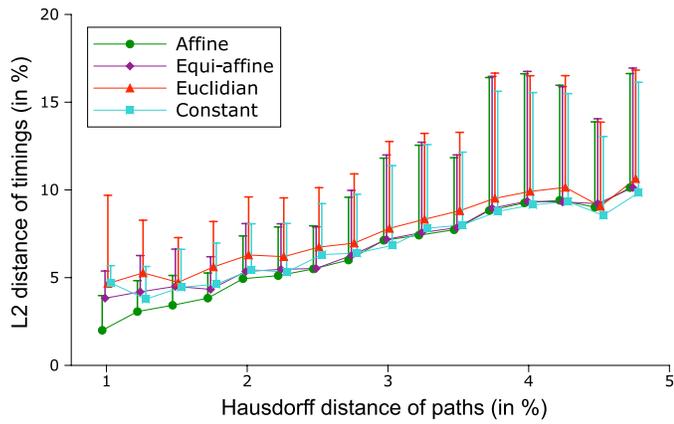

B

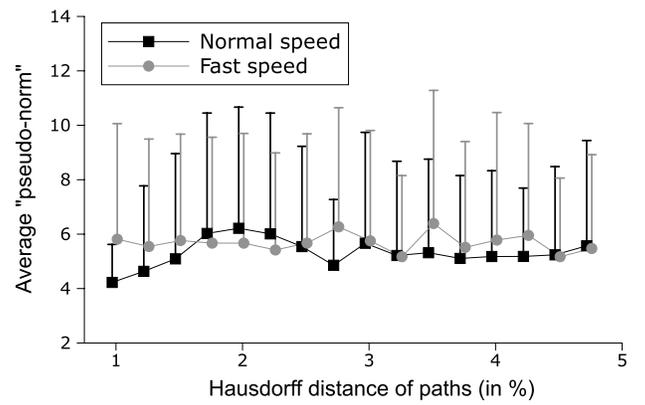

C